\documentclass[a4paper,10pt,fleqn]{report}
\usepackage{epsfig,graphics,graphicx,german,amssymb,amsmath,latexsym,color,psfrag}
\usepackage[authoryear]{natbib}
\usepackage[english]{babel}
\newcommand{\gl}{\;\:=\;\:}

\newcommand{\halb}{\frac{1}{2}}
\DeclareMathOperator{\dilog}{dilog}

\begin{document}
\renewcommand\bibname{{\bf\LARGE References}}
\begin{center}
{\bf\huge A Canonical Measure of Allelic Association} \\
\vspace{1cm}
{\bf Markus Scholz \& Dirk Hasenclever} \\
Institute for Medical Informatics, Statistics and Epidemiology \\
University of Leipzig \\
Haertelstrasse 16-18\\
04107 Leipzig \\
Germany \\
\end{center}

\vspace{1cm}\noindent
{\bf Corresponding author:} \\
Markus Scholz \\
University of Leipzig \\
Institute for Medical Informatics, Statistics and Epidemiology \\
Haertelstrasse 16-18\\
04107 Leipzig \\
Germany \\
Telephone: +49 341 97 16190 \\
Fax: +49 341 97 16109 

\vspace{1cm}\noindent
{\bf Running head:} Canonical measure of association \\

\newpage
\section*{Abstract}
The measurement of biallelic pair-wise association called linkage disequilibrium (LD) is an important 
issue in order to understand the genomic architecture. A large variety of such measures of association in two by two tables have been proposed in the literature.  
\\ 
We propose and justify six biometrical postulates which should be fulfilled by a canonical measure of LD. In short, LD measures are defined as a mapping of two by two probability tables to the set of real numbers. They should be zero in case of independence and extremal if one of the entries approaches zero while the marginals are positively bounded. They should reflect the symmetry group of two by two tables and be invariant under certain transformations of the marginals (selection invariant). There scale should be maximally discriminative for arbitrary tables, i.e. have maximum entropy relative to a calibrating symmetric distribution on the manifold of two by two probability tables. 
\\
None of the established measures fulfil all of these properties in general. We prove that there is a unique canonical measure of LD for each choice of a calibrating symmetric distribution on the set of probability tables. An explicit formula of the canonical measure is derived for Jeffreys' non-informative Dirichlet prior distribution and the uniform Dirichlet distribution. We compare the canonical LD measures with other candidates from the literature. Based on the empirical distribution of association encountered in typical SNP data we recommend the canonical measure derived from Jeffreys' non-informative prior distribution when assessing linkage disequilibrium.
Respective R-procedures are available on request.
\\
In a second part, we consider various estimators for the theoretical LD measures discussed and compare them in an extensive simulation study. The usual plug-in estimators based on frequencies can lead to unreliable estimates. Estimation functions based on the computationally expensive volume measures were proposed recently as a remedy to this well-known problem. We confirm that volume estimators have better expected mean square error than the naive plug-in estimators. But they are outperformed by estimators plugging-in easy to calculate non-informative Bayesian probability estimates into the theoretical formulae for the measures.  
\\ \\ \\ \\
{\bf Keywords:} allelic association, Dirichlet distribution, linkage disequilibrium, maximum entropy, two by two contingency tables

\newpage

\section*{1. Introduction}
\setcounter{chapter}{1}
\subsection*{1.1. Background}

Modern genetic high-through-put methods increasingly provide medium to large size data sets that consist of high dimensional vectors of binary markers. We have been particularly motivated by the example of SNP-chips that address up to one million of biallelic single nucleotide polymorphisms (SNPs). Another example of this data type is patterns of genomic aberration in tumours that can be measured based again on SNP-chip technology or by matrix competitive genome hybridisation (mCGH).  

We restrict ourselves to one sample problems as opposed to two or more sample problems encountered in the context of disease association case-control studies. The focus is to detect highly linked pairs of markers. In the case of SNPs this kind of association is called linkage disequilibrium (LD). Highly linked SNPs are interpreted to be inherited together. LD indicates that a recombination event between the two sites was rare in the population under study. However, there may be other reasons for high LD such as admixture or selection. Linkage has been analysed to understand the genomic architecture especially with respect to recombination hot-spots and jointly inherited haplotype blocks \citep{Schu,Serv}. In the following we always restrict ourselves to LD between two biallelic markers. 

A basic step in analysing such data is assessing associations between markers in a very large number of two by two tables and comparing associations between tables. A bewildering plethora of measures of association are used in the literature \citep{Dev,Hed,T}. Some suggestions on the preferred use of single measures were made \citep{Dev,Mul}. Most of these arguments are based on biological issues such as dependence on allelic frequencies and rate of decay \citep{Hed} or on practical applications such as correlation of test statistics \citep{Prit} and determination of haplotype blocks \citep{Gabi}. 

After a short review of different LD measures, we propose and justify biometrical and statistical postulates to choose between measures of association in the one sample case. We conclude that none of the established LD measures fulfil all of the desirable properties in general. We construct a family of canonical linkage disequilibrium measures which fulfil all of our postulates. Family members differ in the choice of a symmetric Dirichlet distribution on the set of all two by two contingency tables. These Dirichlet distributions calibrate the scale of the measure which essentially measures the extremacy of LD relative to 
the given distribution. The new measures are compared with the established once. Finally, the problem of estimation of the new measure is addressed and different estimators are compared in a simulation study.       

\subsection*{1.2. Measures of Linkage Disequilibrium}
We consider to analyse contingency tables of two biallelic markers at one strand of the genome. Let ${\mathbb T}$ be the manifold of all tetranominal probability models written as a two by two table of probabilities: ${\mathbb T}$ consists of all two by two matrices $t$ with entries $p_{ij}\in {\mathbb R}$, ($i,j\in \{0,1\}$) fulfilling the properties
$p_{ij}>0$, $\sum_{i,j}p_{ij}=1$. The $p_{ij}$ denote the probabilities of the corresponding combination of the 
two alleles of the markers $i$ and $j$. In the following, we abbreviate $\sum_{i=0}^1\sum_{j=0}^1=\sum_{i,j}$, $p_{i.}=p_{i0}+p_{i1}$ and
$p_{.j}=p_{0j}+p_{1j}$ for convenience. Here, the marginals $p_{i.}$ and $p_{.j}$ denote the frequencies of the alleles of the two markers. 
\\
Statistically, a measure of LD is simply a measure of association in the contingency table $t$. The
following measures were defined in literature: \\
{\it D:} The measure $D$ is the absolute deviation of the observation from the expectation that the alleles of marker $i$ are randomly 
combined with alleles of marker $j$ under the assumption of constant marginals. Hence:
\begin{eqnarray*}
D&=&p_{00}-p_{0.}p_{.0}
\end{eqnarray*}
This measure is zero in case of independence of the markers but extremal values depend on the marginals.
\\
{\it Lewontin's $D^\prime$} \citep{Lew}:
The widely used measure $D^\prime$ is a standardisation of the original measure $D$:
\begin{eqnarray*}
D^\prime&=&\frac{D}{D_{max}} \qquad\mbox{where}\quad D_{max}=\left\{\begin{array}{l@{\quad\mbox{if}\quad}l}
\min\left\{p_{0.}p_{.1},p_{.0}p_{1.}\right\} & D\ge 0 \\ 
\min\left\{p_{0.}p_{.0},p_{1.}p_{.1}\right\} & D<0 \end{array} \right.
\end{eqnarray*}
Lewontin's $D^\prime$ ranges from $-1$ to $1$ and tends to these values if one of the $p_{ij}$ tends to zero while the marginals are bounded away from zero. \\
{\it Correlation coefficient $r$} \citep{HR}: The usual correlation coefficient applied to binary data has similar popularity as $D^\prime$. It also ranges from $-1$
to $1$ where an absolute value of $1$ is obtained when a diagonal of $t$ tends to zero:
\begin{eqnarray*}
r&=&\frac{D}{\sqrt{p_{0.}p_{.0}p_{1.}p_{.1}}}
\end{eqnarray*}
{\it Odds ratio $\lambda$} \citep{Ed}:
\begin{eqnarray*}
\lambda&=&\frac{p_{00}p_{11}}{p_{01}p_{10}}
\end{eqnarray*}
The odds ratio is the first quantity which is not directly dependent on $D$ and the marginals. It is well known that $\lambda$ is
independent of selection of single rows or columns of the table $t$. It is thus often used analysing (two sample) case-control studies. The odds ratio is extremal if one of the $p_{ij}$ tends to zero while the marginals are bounded away from zero.\\
{\it Yule's $Q$} \citep{Yule}:
\begin{eqnarray*}
Q&=&\frac{\lambda-1}{\lambda+1}
\end{eqnarray*}
Since the common odds ratio $\lambda$ is not standardised, this quantity has been defined as a function of $\lambda$ 
which is bounded to $[-1,1]$. It can also be written as a difference of the two conditional probabilities $\frac{p_{00}p_{11}}{p_{00}p_{11}+p_{01}p_{10}}$ and
$\frac{p_{01}p_{10}}{p_{00}p_{11}+p_{01}p_{10}}$ \citep{Hart}.\\
{\it Mutual information $M\!I$} \citep{Li,Shan}:
\begin{eqnarray*}
M\!I&=&\sum_{i,j} p_{ij} \log_2 p_{ij} - \sum_{i=0}^1 p_{i.}\log_2 p_{i.} - \sum_{j=0}^1 p_{.j}\log_2 p_{.j} 
\end{eqnarray*}
$M\!I$ has its minimal value zero when $p_{00}=p_{0.}p_{.0}$ and its 
maximal value one only if $t$ is diagonal with either $p_{00}=p_{11}=\frac{1}{2}$ or $p_{01}=p_{10}=\frac{1}{2}$. Hence, $M\!I$ is not normalized.
\\
A further measure $Dvol$ has been proposed by {\sc Chen et al.} \citep{Chen}. Rather than 
a new measure of LD it is an alternative estimation function for $D^\prime$. $Dvol$ is defined as the fraction of 
contingency tables with less extreme $D$ in the space of all contingency tables with same sign of $D$ and fixed marginals. \\
The use of these measures has been discussed extensively and it has been recommended to calculate $r$ when one marker is used to predict another marker and to use $D^\prime$ as a measure of recombination probability \citep{Dev,Mul}. However,
all these recommendations lack of a clear definition of desired statistical properties of a measure of LD. Hence, the use and interpretation of these measures remain vague.  

\section*{2. The canonical measure of Linkage Disequilibrium}
\setcounter{chapter}{2}
\setcounter{equation}{0}

\subsection*{2.1. Postulates for a Canonical Measure of Linkage Disequilibrium}
In this section we list postulates for a canonical measure of LD giving both biological and mathematical justifications. 
\begin{description}
\item[P1 (Domain of association measure):] A measure of LD is a continuous function $\eta: \mathbb{T} \rightarrow \mathbb{R}$.
\end{description}
The LD measure $\eta$ is formally defined on the manifold $\mathbb{T}$ of tetranomial probability models (with two by two lay-out), not on a set of concrete realisations (for example as two by two data tables of sample size $N$). Defining the LD measure and estimating it from concrete data are radically separate tasks.
\begin{description}
\item[P2 (Lack of association and complete linkage disequilibrium):] Tables in linkage equilibrium show no association that is they fulfil $p_{ij}=p_{i.}p_{.j}$ $(i,j \in \{0,1\})$. Complete LD is present whenever \textbf{at least one} $p_{ij}$ approaches zero while the marginals retain a positive lower limit. 
\end{description}
When a new SNP emerges in a population by a single mutation event, the new allele is exclusively found in conjunction with only one of the two alleles of an already existing SNP. As long as no recombination event occurs, the new SNP remains in complete LD with the other SNP. The corresponding two by two table features a single zero cell. The measure $r$ discussed above becomes extremal only if the underlying table approaches diagonal form. A diagonal structure suggests population admixture, specific negative selection or a mutation event giving rise to both SNPs simultaneously. The postulate covers all these cases.
\begin{description}  
\item[P3 (Symmetry):] The LD measure $\eta$ reflects the two by two symmetry structure of $\mathbb{T}$: \\ 
a) $\eta$ is invariant under permutation of the SNPs which means matrix transposition. \\
b) $\eta$ changes sign when alleles of a SNP, which are rows or columns of the table, are transposed.
\end{description}
Since we consider the one-sample case none of the binary markers is distinguished. SNP transposition and allele transposition generate a symmetry group (a dihedral group $D_{4}$)
that should leave a measure of association essentially invariant.
\\Requiring antisymmetry and thus introducing a sign to the linkage disequilibrium measure is convenient in applications where we have well defined marked states of the markers \citep{T}. Alternatively, one may just take the absolute value of $\eta$ to obtain a measure invariant under the full symmetry group. 
\\ Note that P3 implies that tables in linkage equilibrium are mapped to zero, since transposing preserves the condition of no association. 
\begin{description}
\item[P4 (Selection invariance:)] The LD measure $\eta$ is not changed by selection of alleles of one SNP that does not affect the corresponding allele frequency ratios of the other SNP. In other words, multiplication of columns or rows by positive numbers and renormalisation of the table does not change the measure of association.
\end{description}
Selection of alleles and genetic drift is common during the course of evolution in a population. Allele frequencies fluctuate and differ in distinct populations. The measure of LD should capture an intrinsic property of the genome architecture that reflects the structural probability of a recombination event between two markers. It should thus not depend on the marginal allelic frequencies that happen to be observable in a given population. \\
In addition, there may occur sampling selection in obtaining the data that introduces bias. Using a selection invariant measure of association safeguards against this danger up to a certain degree.\\
Selection invariance is particularly important if the measure of association is intended to be meaningfully compared between tables with markedly different marginal distributions (allele frequencies).
\begin{description}
\item[P5 (Standardization):] The LD measure $\eta$ is standardized to values in $(-1,1)$. The extremal values $\pm1$ stand for perfect LD.  To achieve uniqueness we require that $\eta(t) \rightarrow 1 $ for $t \rightarrow \left( {0.5 \atop 0}{0 \atop 0.5}\right)$
\end{description}
\begin{description}
\item[P6 (Maximum Entropy):] The LD measure $\eta$ should classify arbitrary tables of $\mathbb{T}$ 
as discriminatively as possible. To formalise the notion of arbitrary tables we choose a 
symmetric ''non-informative'' distribution on $\mathbb{T}$ and require that the induced 
distribution of the LD measure $\eta$ has maximum entropy. That is it is uniform on $(-1,1)$.
\end{description}
Postulates 1 to 5 do not determine yet the scale for association values between $0$ (independence) and the extremal values $\pm1$. The sixth postulate requires using a most informative and discriminative scale. The LD measure $\eta$ should be a good general classifier of arbitrary tables of $\mathbb{T}$. No particular value of the LD measure $\eta$ within $(-1,1)$ should be privileged. Therefore, we require that $\eta$ have a uniform (maximum entropy) distribution on $(-1,1)$ when sampling arbitrary tables from $\mathbb{T}$. 

Postulate 6 implies the choice of a symmetric calibrating distribution on $\mathbb{T}$ to specify sampling ''arbitrary'' tables. In theorem 2 we will see that $\eta$ has an intuitive interpretation based on the proportion of tables having less extreme LD than the given one in the chosen distribution.  
\\Later we calculate the LD measure $\eta$ for Dirichlet distributions 
${\cal D}\left(\alpha\right)$, $\alpha=\left( \alpha_{00}, \alpha_{01}, \alpha_{10}, \alpha_{11}\right)$ ($\alpha_{ij}>0$). 
This family of distributions is often used in a Bayesian context as prior distribution for contingency tables since it is the conjugate prior to the multinomial distribution \citep{Geis,Wal}. The density of the Dirichlet distribution is given by:
\begin{eqnarray*}
f_{{\cal D}\left(\alpha\right)}&=&\frac{1}{B\left(\alpha\right)}\prod_{i,j}p_{ij}^{\alpha_{ij}-1}
\quad\mbox{where}\qquad
B\left(\alpha\right)\gl\frac{\prod_{i,j}\Gamma\left(\alpha_{ij}\right)}{\Gamma\left(\sum_{i,j}\alpha_{ij}\right)}
\end{eqnarray*}
is the Beta-function and $\Gamma$ denotes the Gamma function. 
\\
The Dirichlet distributions are symmetric (invariant under transpositions of columns or rows) if and only if all $\alpha_{ij}$ are equal. 
For simplicity of notation, we identify the vector $\alpha$ with one of its components in the symmetric case.
\\
A principled choice for a symmetric non-informative distribution on $\mathbb{T}$ to 
define the canonical LD measure $\eta$ may be the well-known non-informative Jeffreys' prior $\alpha=\frac{1}{2}$ 
\citep{Geis,Jef}. In addition, with ${\cal D}\left(\halb\right)$ the distribution of the minor marginal frequencies is uniform on $[0;0.5]$ which in our experience is often encountered in SNP-array-data. We also discuss the choices $\alpha=1$ and $\alpha=2$.

\subsection*{2.2. Construction of a Canonical Measure of Linkage Disequilibrium}
We will now explore the consequences of this set of postulates. In particular, we will show that for any continuous symmetric distribution on $\mathbb{T}$ the canonical LD measure $\eta$ exists and is unique. Later we will calculate $\eta$ for symmetric Dirichlet distributions. We start with a closer look at selection invariance.  
\\
Selection invariance may be formalised as the action of a suitable group $G$ on $\mathbb{T}$: Consider the group $G=(\mathbb{R^+}\times\mathbb{R^+},\cdot)$ with component-wise multiplication. \\
For every $(\mu,\nu) \in \mathbb{R^+}\times\mathbb{R^+}$ we define a map: $g(\mu,\nu):\mathbb{T}\longrightarrow \mathbb{T}$
\begin{eqnarray} 
t=\left({p_{00}\atop p_{10}}{p_{01}\atop p_{11}}\right) \longmapsto g(\mu,\nu)(t)&=&\frac{1}{\mu\nu p_{00}+\mu p_{01}+\nu p_{10}+p_{11}}\left(\begin{array}{ll} \mu\nu p_{00} & \mu p_{01} \\
\ \nu p_{10} &  p_{11} \end{array}\right) 
\end{eqnarray}
Since $g(\mu,\nu) \circ g(\mu^\prime,\nu^\prime) = g(\mu\cdot\mu^\prime,\nu\cdot\nu^\prime)$ and $g(1,1)=Id_\mathbb{T}$ this defines a G-group action on $\mathbb{T}$. 
A function $\eta: \mathbb{T} \rightarrow \mathbb{R}$ is defined as selection invariant if $\eta(t)=\eta(g(\mu,\nu)(t))$ for all $(\mu,\nu) \in \mathbb{R^+}\times\mathbb{R^+}$.
Lying in the same group orbit defines an equivalence relation on $\mathbb{T}$: We say  two elements $t_1,t_2 \in \mathbb{T}$ are equivalent $t_1 \sim t_2$ if and only if there are $(\mu,\nu)\in\mathbb{R^+}\times\mathbb{R^+}$ with $g(\mu,\nu)(t_1)=t_2$. 
Thus every selection invariant function $\eta: \mathbb{T} \rightarrow \mathbb{R}$ induces a well-defined map $\tilde{\eta}: \tilde{\mathbb{T}}  \rightarrow \mathbb{R}$ on the quotient space $\tilde{\mathbb{T}}=\mathbb{T} /\sim$. \\ \\
{\bf Theorem 1 (odds ratio):}\\
a) The odds ratio $\lambda:\mathbb{T} \rightarrow \mathbb{R}; \ t=\left({p_{00}\atop p_{10}}{p_{01}\atop p_{11}}\right) \mapsto \lambda(t)=\frac{p_{00}p_{11}}{p_{01}p_{10}}$ is selection invariant.\\ 
b) The odds ratio induces a homeomorphism  $\tilde{\lambda}: \tilde{\mathbb{T}}\stackrel{~}{\rightarrow}\mathbb{R^+}$.\\
c) The inverse mapping $\tilde{\lambda}^{-1}: \mathbb{R^+} \rightarrow \tilde{\mathbb{T}}$ can be described by 
$l  \longmapsto \left[\left(\begin{array}{ll} \frac{\sqrt{l}}{2\cdot(1+\sqrt{l})} & \frac{1}{2\cdot(1+\sqrt{l})} \\
\frac{1}{2\cdot(1+\sqrt{l})} &  \frac{\sqrt{l}}{2\cdot(1+\sqrt{\l})} \end{array}\right)\right]$\\
d) Every selection-invariant function $\eta: \mathbb{T} \rightarrow \mathbb{R}$ can be written as a function of $\lambda$, namely $\eta=(\tilde{\eta}\circ\tilde{\lambda}^{-1})\circ\lambda$.
\\ \\
{\bf Proof:} a) is easily verified. Every equivalence class [t] in $\tilde{\mathbb{T}}$ has a representant with marginals $\frac{1}{2}$, namely $[g(\sqrt{\frac{p_{11}p_{10}}{p_{00}p_{01}}},\sqrt{\frac{p_{11}p_{01}}{p_{00}p_{10}}})(t)]$ which has the form given in c). d) is trivial. \hfill $\Box$
\\ \\
For every distribution ${\cal D}$ on $\mathbb{T}$ the odds ratio $\lambda$ induces the distribution $\lambda_*({\cal D})$ of the corresponding odds ratios on $\mathbb{R^+}$. 
\\ \\
{\bf Theorem 2 (Existence of a canonical LD measure):} Let ${\cal D}$ be a symmetric (non\--infor\-mative) distribution on $\mathbb{T}$. Let $L$ denote the cumulative distribution function of $\lambda_*({\cal D})$ on $\mathbb{R^+}$.\\
Define 
\begin{eqnarray}
\eta(t)&=&2L\left(\lambda(t)\right)-1
\end{eqnarray}
Then $\eta$ is the unique canonical LD measure that fulfils postulates P1-P6 chosen ${\cal D}$.
\\ \\
{\bf Proof:} Because of theorem 1 the measure $\eta$ depends only on the odds ratio $\lambda$
of the table. By construction $\eta$ is uniform on $(-1,1)$. The remaining postulates and uniqueness are easily verified. \hfill $\Box$
\\ \\
{\bf Remark:} Note that this construction provides $\eta$ with an intuitive interpretation. $\eta(t)$ is a signed measure  of extremality of the odds ratio $\lambda(t)$ relative to an underlying calibrating distribution ${\cal D}$ of arbitrary tables on $\mathbb{T}$. It is based on the proportion of tables with less extreme odds ratio in this distribution.

\subsection*{2.3. Calculation of the Canonical Measure of Linkage Disequilibrium for Symmetric Dirichlet Distributions}
We determine the distribution function of the odds ratio $\lambda$ under the Dirichlet distribution
${\cal D}(\alpha)$. Define
\begin{eqnarray*}
\Omega_{\Lambda}&=&\left\{ t\in\mathbb{T} \quad : \quad \lambda (t)<\Lambda \right\} \quad \forall \Lambda \in \mathbb{R}^+
\end{eqnarray*}
We calculate 
\begin{eqnarray}
\int_{\Omega_\Lambda} f_{{\cal D}(\alpha)} &=& \frac{1}{B\left( \alpha\right)}
\int_{\Omega_\Lambda} p_{00}^{\alpha_{00}-1}p_{01}^{\alpha_{01}-1}p_{10}^{\alpha_{10}-1}
\left(1-p_{00}-p_{01}-p_{10}\right)^{\alpha_{11}-1} \, dp_{00} \, dp_{01} \, dp_{10}
\end{eqnarray} 
Using the formula
\begin{eqnarray*}
p_{10}&=&\left( 1-p_{00}-p_{01}\right)\frac{p_{00}}{\lambda p_{01} + p_{00}} 
\end{eqnarray*}
we transform the coordinates $\left( p_{00} ,p_{01}, p_{10} \right)$ to
$\left( p_{00} ,p_{01}, \lambda \right)$ with the corresponding functional determinant 
\begin{eqnarray*}
\left| \frac{\partial\left(p_{00} ,p_{01}, p_{10} \right)}{\partial \left( p_{00} ,p_{01} ,\lambda\right)} \right|
&=&\left( 1-p_{00}-p_{01}\right) \frac{p_{00} p_{01}}{\left( \lambda p_{01} + p_{00} \right)^2}
\end{eqnarray*} 
Furthermore, $\Omega_{\Lambda}$ can be parameterized by $p_{00}\in \left( 0,1\right)$,
$p_{01} \in \left( 0,1-p_{00} \right)$ and $\lambda \in \left( 0, \Lambda\right)$. Hence,
(2.3) can be written as
\begin{eqnarray}
\int_{\Omega_\Lambda} f_{{\cal D}(\alpha)} &=&\int_0^\Lambda l(\lambda) \, d\lambda
\end{eqnarray}
where
\begin{eqnarray}
l\left(\lambda\right) &=&
\frac{\lambda^{\alpha_{11}-1}}{B(\alpha)}\int_0^1 \! \int_0^{1-p_{00}} \frac{p_{00}^{\alpha_{00}+\alpha_{10}-1} p_{01}^{\alpha_{01}+\alpha_{11}-1}
\left( 1-p_{00}-p_{01} \right)^{\alpha_{10}+\alpha_{11}-1}}
{\left( \lambda p_{01}+p_{00}\right) ^{\alpha_{10}+\alpha_{11}}} \, dp_{01} \,dp_{00}
\end{eqnarray}
is the probability density of $\lambda$ under the Dirichlet distribution ${\cal D}\left( \alpha \right)$. 
\\
A plot of this density can be found in figure 1 for some special Dirichlet distributions. 

In the following, we consider symmetric Dirichlet distributions and denote the corresponding 
canonical LD measure with $\eta_\alpha$.  

\noindent
In case of $\alpha\in\left\{\halb,1\right\}$ there are analytic formulae for $\eta_\alpha$ which will be derived now.
\\ \\
{\bf Theorem 3 (Analytic formula for $\eta_1$):} 
\begin{eqnarray}
\eta_1\left(\lambda\right)&=& 2\frac{\lambda^2-\lambda-\lambda\ln\lambda}{\left(\lambda-1\right)^2}-1  \qquad\forall\lambda\not= 1
\end{eqnarray} 
The gap of definition at $\lambda=1$ can be removed by taking the limit $\lim_{\lambda\to 1}
\eta_1\left(\lambda\right)=0$.
\\ \\
{\bf Proof:}
At first we solve the double integral for $l$ in (2.5)
\begin{eqnarray}
l\left(\lambda\right) &=&6 \int_0^1 \! \int_0^{1-p_{00}} \frac{p_{00}p_{01}\left(1-p_{00}-p_{01}\right)}
{\left( \lambda p_{01}+p_{00}\right)^2} \, dp_{01} \,dp_{00}
\end{eqnarray}
Using partial fraction decomposition we obtain after some calculation
\begin{eqnarray}
l\left(\lambda\right) &=&-6\int_0^1 \frac{p_{00}}{\lambda^3} \left\{
2\lambda-2\lambda p_{00} + \left(2p_{00} + \lambda - \lambda p_{00}\right) 
\ln \frac{p_{00}}{p_{00}+\lambda-\lambda p_{00}} \right\} \, dp_{00}
\end{eqnarray}
The only summand in (2.8) causing difficulties is $\int p_{00} \left(2p_{00} + \lambda - \lambda p_{00}\right)
\ln \left( p_{00}+\lambda-\lambda p_{00} \right) \, dp_{00}$. It can be solved by 
substitution of $p_{00} \left (1-\lambda\right)$, where it is necessary to distinguish $\lambda>1$
and $\lambda<1$. A singularity occurs for $\lambda=1$ which can be removed separately.
After a longer calculation it follows that
\begin{eqnarray}
l\left(\lambda\right) &=& \frac{2-2\lambda+\ln\lambda+\lambda\ln\lambda}
{\left( \lambda-1\right)^3} \qquad\forall\lambda\not= 1   
\end{eqnarray}
and $l(1) = \lim_{\lambda\to 1} l(\lambda) = \frac{1}{6}$.

With equation (2.9) the third integral can be calculated
\begin{eqnarray*}
\int_0^\Lambda l(\lambda )\,d\lambda &=&
\int_0^\Lambda 2\frac{1-\lambda+2\ln\lambda}{\left(\lambda-1\right)^3} + \frac{\ln\lambda}{\left(\lambda-1\right)^2} \, d\lambda
\end{eqnarray*}
Both summands can be dealt with using partial integration 
\begin{eqnarray*}
\int_0^\Lambda l(\lambda )\, d\lambda &=&\frac{\Lambda^2-\Lambda-\Lambda\ln\Lambda}{\left(\Lambda-1\right)^2}
\end{eqnarray*}
\hfill $\Box$
\noindent
\\
{\bf Theorem 4 (Analytic formula for $\eta_\halb$):} 
\begin{eqnarray}
\eta_\halb\left(\lambda\right)&=& \frac{2}{\pi^2} \int_0^\lambda \frac{\ln y}{\sqrt{y}\left(y-1\right)} \,dy -1
\end{eqnarray} 
This integral can also be expressed in terms of the dilogarithm function $\dilog x =-\int_0^x\frac{\ln 
\left| 1-y\right|}{y} \, dy$ (see \citet{Max} for properties of $\dilog$) for which good numerical procedures are available \citep{Koel}.
\begin{eqnarray}
\eta_\halb\left(\lambda\right) &=& \frac{4}{\pi^2} \left\{\ln\left(\sqrt{\lambda}\right)\ln\left|\frac{\sqrt{\lambda}-1}{\sqrt{\lambda}+1}\right|+\dilog\left(\sqrt\lambda\right)
-\dilog\left(-\sqrt{\lambda}\right)\right\}-1
\end{eqnarray}
Again, there is a gap of definition at $\lambda=1$ which can be resolved by $\lim_{\lambda\to 1}
\eta_\halb\left(\lambda\right)=0$.
\\ \\
{\bf Proof:} Equation (2.10) follows directly from equation (2.5) after elementary integration. Equation (2.11) follows from (2.10) after substitution of $\sqrt{y}$ and partial integration. \hfill $\Box$

\subsection*{2.4. Comparison of the Canonical LD Measures with Commonly Used LD Measures}
In this section we compare the canonical LD measures, especially $\eta_1$ and $\eta_\halb$, with established measures of LD represented by its most common
represantatives $D^\prime$, $r$ and $Q$. We analyse their commonalities, differences and behaviour in special situations with the help of seven remarks.  \\ \\
{\bf Remark 1:} All LD measures based on the odds ratio $\lambda$, such as $\eta_\alpha$ (for $\alpha>0$) and $Q$, are strictly monotone functions of each other. Figures 2 illustrates these functional relations. 
\\ \\
{\bf Remark 2:} The measure $Q$ is a good approximation for $\eta_2$ with $\max_{t\in{\mathbb T}} \left|Q\left(t\right)-\eta_2\left(t\right)\right|\approx 0.035$. Hence, $Q$ is approximately uniformly distributed under ${\cal D}\left(2\right)$. Optimal agreement of $Q$ and $\eta_\alpha$ is obtained for $\alpha\approx 1.77$ with $\max_{t\in{\mathbb T}} \left|Q\left(t\right)-\eta_\alpha\left(t\right)\right|\approx 0.013$.
\\ \\
{\bf Remark 3:} The LD measure $r$ needs a diagonal structure to approach the extremal values of $\pm1$ while $D^\prime, Q$ and $\eta_\alpha$ (for $\alpha>0$) could become extremal whenever one table entry tends to zero. Thus for tables with  $D', Q$ or $\eta_\alpha$ near 1,  $r$ may assume values in $(0,1)$. Compare Figure 3.
\\ \\
{\bf Remark 4:} The LD measures $D^\prime$ and $r$ are not selection invariant. 
\\ \\
{\bf Proof:} We prove this by an example. Let $t_1=\frac{1}{8}\left( {3 \atop 1}{1 \atop 3}\right)$
and $t_2=\frac{1}{12}\left( {9 \atop 1}{1 \atop 1}\right)$. It holds that $t_1\sim t_2$ with $\mu=\nu=3$. 
One calculates that $D^\prime \left( t_1\right) = r\left( t_1\right)=\frac{1}{2}$
while $D^\prime\left( t_2\right) = r\left( t_2\right)=\frac{2}{5}$. On the other hand $\lambda\left( t_1\right)=\lambda\left( t_2\right)=9$. \hfill $\Box$
\\ 
It is not surprising that measures depending on $D$ are not selection-invariant since the concept of $D$ is based on constant marginals while selection may change these.
\\\\
{\bf Remark 5:} For every $\varepsilon>0$ there is a table $t_{\varepsilon}\in\mathbb{T}$ such that $\left| D^\prime\left( t_{\varepsilon}\right) \right|<\varepsilon$ and $\lambda\left(t_{\varepsilon}\right)>\frac{1}{\varepsilon}$. Hence, there are tables for which $D^\prime$ measures almost no LD and $\lambda$, $Q$ and $\eta_\alpha$ measure almost perfect LD. \\ \\
{\bf Proof:}
Consider for example
\begin{eqnarray*}
t_{\delta}&=&
\left( \begin{array}{ll} \left(1-\delta\right)^2+\sqrt{\delta^3} & 
\delta\left(1-\delta\right)-\sqrt{\delta^3} \\ \delta\left(1-\delta\right)-\sqrt{\delta^3} & 
\delta^2+\sqrt{\delta^3} \end{array}\right) 
\end{eqnarray*}
for $\delta>0$ sufficiently small. Then it 
follows that $\lim_{\delta\to 0} D^\prime\left(t_\delta\right)=0$ and $\lim_{\delta\to 0} \lambda\left(t_\delta\right)=\infty$. \hfill $\Box$
\\ 
Note however that this abnormality only occurs when three table entries tend to zero. Compare Figure 4.
\\ \\
{\bf Remark 6:} For tables with not too imbalanced marginals, monotonicity between $D'$ and odds ratio based measures is essentially preserved. Table 1 shows the Kendall correlation coefficient determined for tables 
with specified marginals. 
\\ \\
{\bf Remark 7:} On a set of tables with constant marginals, $p_{00}$  is uniformly distributed under ${\cal D}\left(1\right)$. Thus after standardisation with $D_{max}$, $D^\prime$ is uniformly distributed on $\left( -1,1 \right)$ and has maximum entropy in this case. By construction the canonical LD measures $\eta_1$ and $\eta_\halb$ are uniformly distributed when paired with their respective Dirichlet distribution. Compare Figure 5.

\section*{3. The Estimation Problem for LD Measures}
\setcounter{chapter}{3}
\setcounter{equation}{0}

We now investigate various estimators for the theoretical linkage disequilibrium measures discussed above. Here, we do not address the problem that the entries $n_{ij}$ sometimes must also be estimated from real data by a phasing algorithm 
in case of double heterozygote markers. This can be done for example with the help of the exact solution 
of an EM-algorithm (see \citet{Weir} for details). At this step it is necessary to assume haploid populations or polyploid populations in Hardy-Weinberg equilibrium.  
\noindent

Estimation is based on observed contingency tables 
$t_N = \left( {n_{00} \atop n_{10}}{n_{01} \atop n_{11}}\right)$
with $n_{ij} \in \mathbb{N}$ and $\sum_{i,j} n_{ij} = N$. The table $t_N$ 
is regarded as a random realization of the true table $t \in \mathbb{T}$ under the corresponding tetranomial distribution with sample size $N$. The tables $t_N$ form a sample space $\mathbb{T}_{N}$ of all possible realizations of $t$ after $N$-fold sampling. 
In the following, we will define and compare different families of consistent estimators for linkage disequilibrium measures given an observation
$t_N\in{\mathbb T}_{N}$. 

\subsection*{3.1. Estimators for LD Measures}
The common approach is the use of ''plug-in'' estimates, where the probability estimates 
$\hat{p}_{ij}$ of $p_{ij}$ are inserted into the theoretical formula of a measure. 
Often, the frequency maximum-likelihood estimates $\frac{n_{ij}}{N}$ of $p_{ij}$ are used. 
However this may lead to inflated or undefined estimates of the desired quantity especially 
in case of small sample sizes \citep{Tear}. This approach has been used both extensively and 
carelessly in the literature to estimate for example $D^\prime$ and $r$. We will denote 
corresponding estimators as {\it naive estimators} $N\!E$.\\
For any LD measure $M $ it reads
\begin{eqnarray}
\hat{M}^{N\!E}\! \left(t\right)&=& M\!\left(\hat{t}\right) \qquad\mbox{with}\quad
\hat{t}=\left( \begin{array}{ll} \frac{n_{00}}{N} & \frac{n_{01}}{N} \\ \frac{n_{10}}{N} & \frac{n_{11}}{N}
\end{array}\right)
\end{eqnarray}
An alternative approach is using ''non-informative'' Bayesian probability estimates for $p_{ij}$ \citep{Wal}.
\begin{eqnarray}
\tilde{p}_{ij}&=&\frac{\alpha+n_{ij}}{4\alpha + N} \qquad\mbox{with}\quad \alpha\in\left\{\halb,1\right\}
\end{eqnarray}
which is the expectation of the posteriori distribution ${\cal D}\left(\alpha+n\right)$ were 
$n=\left(n_{00},n_{01},n_{10},n_{11}\right)$.
Calculating $M$ with the help of $\tilde{p}_{ij}$ instead of $p_{ij}$ yields 
a consistent {\it semi-naive estimator} $S\!N\!E$. It has the form
\begin{eqnarray}
\hat{M}^{S\!N\!E}\! \left(t\right) &=&
M\! \left(\tilde{t}\right) \qquad\mbox{with}\quad
\tilde{t}=\left( \begin{array}{ll} \tilde{p}_{00} & \tilde{p}_{01} \\ \tilde{p}_{10} & \tilde{p}_{11}
\end{array}\right)
\end{eqnarray}
A further approach is to calculate the expectation of $M$ under a posteriori
distribution which is the ordinary {\it Bayes estimator} $B\!E$
\begin{eqnarray}
\lefteqn{ \hat{M}^{B\!E} \left(t\right) \gl
\int_{\mathbb{T}} f_{{\cal D}\left( \alpha+n \right)} 
M\left(t\right) \, dt } \nonumber\\
&&\gl\frac{1}{B\left(\alpha+n\right)}
\int_0^1 \!\! \int_0^{1-p_{00}} \!\! \int_0^{1-p_{00}-p_{01}}
p_{00}^{\alpha_{00}+n_{00}-1} p_{01}^{\alpha_{01}+n_{01}-1} p_{10}^{\alpha_{10}+n_{10}-1}\cdot \\
&& 
\quad\quad\left( 1-p_{00}-p_{01}-p_{10} \right)^{\alpha_{11}+n_{11}-1} 
M\left(t\right) \, dp_{00} \, dp_{01} \, dp_{10} \nonumber
\end{eqnarray}
\noindent
Finally, some LD measures can be estimated with a so-called volume formula. The concept of 
volume measures can be traced back to \citet{Hot} and has been applied to 
contingency tables by \citet{DE} and to linkage disequilibrium measures by 
\citet{Chen}. In the simplest case, the idea is to count the number of tables 
which are less ''extreme'' than an observed table and compare this ''volume'' with the total 
volume of all possible tables. \\
\citet{Chen} defined $Dvol$ as the number of tables with fixed marginals, fixed sign of $D$ and less extreme  
values of $D$ divided by the number of tables with fixed 
marginals and fixed sign of $D$. In the original definition, this measure is always greater 
or equal to 0 and less than 1.  For better comparability with the other estimators, we consider 
an obvious signed version $\hat{D}^\prime_{V\!E}$ with values in the interval $\left(-1,1\right)$
by assigning the sign of $D$ of the observed table $t_N$. The claimed advantage of this 
estimating procedure is that $\hat{D}^\prime_{N\!E}$ is inflated \citep{Tear} while $\hat{D}^\prime_{V\!E}$ is not in case of tables with small numbers where the occurrence of zeros is likely \citep{Chen}. \\
Generalising this approach, we define an estimation function $\hat{\eta}_\alpha^{V\!E}$ for 
$\eta_\alpha$. 
Any Dirichlet distribution ${\cal D}\left(\alpha\right)$ induces a probability distribution 
$w_\alpha$ on the sample space ${\mathbb T}_{N}$ and thus a discrete probability distribution of the 
corresponding odds ratios $\lambda\left(\tilde{t}\right)$.
We define $\hat{\eta}_\alpha^{V\!E}\left( t_N \right)$ in analogy to equation (2.2) as the 
probability under ${\cal D}\left(\alpha\right)$ to obtain an odds ratio less extreme than 
$\lambda\left(\tilde{t}_N \right)$ standardised to $\left( -1,1\right)$. By construction, the 
function $L$ in (2.2) can be interpreted as the ''volume'' of tables with smaller odds ratio than 
$\Lambda$ divided by the total volume of all tables in ${\mathbb T}_N$. 

Note that ${\mathbb T}_N$ contains all tables with fixed sum of entries but not fixed marginals as 
in the definition of $\hat{D}^\prime_{V\!E}$. 
\\
This definition can be used to construct another estimation 
function for $\eta$ directly from the observed contingency table $t_N$. 
In contrast to the construction of $\hat{D}^\prime_{V\!E}$ we will not assume that all alternative tables $t_N \in {\mathbb T}_N$ 
are equally likely. Therefore, we calculate the probability
of a single table $t_N$ given a Dirichlet distribution.
\\ \\
{\bf Theorem 5:} Let $w_\alpha\left(t_N\right)$ be the probability of the table $t_N \in {\mathbb T}_N$ under 
the distribution ${\cal D}\left(\alpha\right)$ on ${\mathbb T}_N$, then we have  
\begin{eqnarray}
w_\alpha\left(t_N\right) &=& N\frac{B\left(N,\sum_{i,j}\alpha_{ij}\right)}{\prod_{i,j} n_{ij} B\left(n_{ij},\alpha_{ij}\right)}
\end{eqnarray}
where we define $n B\left(n,x\right)=1$ for $n=0$, $x>0$ and $B$ is the Beta-function.
\\ \\
{\bf Proof:}
Let ${\cal M}\left(n,p\right)$ be the multinomial distribution of $n_{ij}$ under the probabilities
$p_{ij}$, then  
\begin{eqnarray*}
w_\alpha\left(t_N\right) &=&\mbox{prob}\left(t_N \quad | \quad {\cal D}\left(\alpha\right)\right) \\
&=& \int_{\mathbb T} {\cal M} \left(n,p\right) f_{{\cal D} \left(\alpha\right)} \, d p \\
&=& C\int_0^1 \int_0^{1-p_{00}} \int_0^{1-p_{00}-p_{01}} p_{00}^{\tilde{n}_{00}} p_{01}^{\tilde{n}_{01}}
p_{10}^{\tilde{n}_{10}} \left( 1-p_{00}-p_{01}-p_{10} \right)^{\tilde{n}_{11}} \, dp_{10} dp_{01} dp_{00}
\end{eqnarray*}
where $C=\frac{1}{B\left(\alpha\right)} {N \choose n_{ij}}$ and $\tilde{n}_{ij}=n_{ij}+\alpha_{ij}-1$. Using the identity
\begin{eqnarray*}
\int_0^y x^n \left( y-x\right)^m \,dx &=& y^{m+n+1} B\left(m+1,n+1\right)
\end{eqnarray*}
it follows that
\begin{eqnarray*}
w_\alpha\left(t_N\right) &=&C B\left(\tilde{n}_{10}+1,\tilde{n}_{11}+1\right) B\left(\tilde{n}_{01}+1,\tilde{n}_{10}+
\tilde{n}_{11}+2\right) B\left(\tilde{n}_{00}+1,\tilde{n}_{10}+\tilde{n}_{01}+\tilde{n}_{11}+3\right) \\
&=&C\frac{\prod_{i,j}\Gamma\left(n_{ij}+\alpha_{ij}\right)}{\Gamma\left(\sum_{i,j} n_{ij}+\alpha_{ij}\right)}
\end{eqnarray*}
Rewriting the last equation yields equation (3.5). \hfill $\Box$
\\ \\
{\bf Remark:} Obviously, for the Dirichlet distribution ${\cal D}\left(1\right)$ we have $w=\frac{1}{{N+3 \choose 3}}$ constant.
\\ \\
With the last theorem, the volume estimator for $\eta$ can be defined. Let 
\begin{eqnarray}
\hat{\lambda}\left(t_N\right) &=& \frac{\left(n_{00}+\alpha_{00}\right)\left(n_{11}+\alpha_{11}\right)}
{\left(n_{01}+\alpha_{01}\right)\left(n_{10}+\alpha_{10}\right)}
\end{eqnarray}
be the semi-naive estimator for the odds ratio, then, we define
\begin{eqnarray*}
\hat{l} \left(\lambda\left(t\right)\right)&=&
\sum_{i=0}^N \sum_{j=0}^{N-i} \sum_{k=0}^{N-i-j} w\left(i,j,k,N-i-j-k\right) \chi\left( \hat{\lambda}\left(t_{i,j,k,N-i-j-k}\right),\hat{\lambda}\left(t_N\right)\right) \\
&&\mbox{with}\quad t_{i,j,k,l}= {i \, j \choose k\,l}
\end{eqnarray*}
where the indicator function $\chi$ has the form
\begin{eqnarray*}
\chi\left(\lambda_1,\lambda_2\right) &=& \left\{ \begin{array}{r@{\quad:\quad}l} 1 
& \lambda_1<\lambda_2 \\
\halb & \lambda_1=\lambda_2 \\
0 & \mbox{else} \end{array} \right.
\end{eqnarray*} 
And finally
\begin{eqnarray}
\hat{\eta}^{V\!E}_{{\cal D}\left(\alpha\right)} \left(\lambda\left(t\right)\right)&=&
2\hat{l} \left(\lambda\left(t\right)\right)-1
\end{eqnarray}

\subsection*{3.2. Comparison of Estimators}
We compared these estimation functions in a simulation study. First, we simulate true tables by random drawing 
from specified Dirichlet distributions. From it, true values of the linkage disequilibrium measures can be 
calculated. In the next step, we construct a concrete realization of the true tables by random 
drawing from the corresponding multinomial distribution with different sample sizes $N$. 
The estimation functions are compared with respect to their expected mean square error. 

The analysis is performed for $\eta_1$, $\eta_\halb$, $D^\prime$, $r$, $Q$ and corresponding estimation functions.
Results can be found in table 2. 

Because of different variances of the true measures, we can only compare different estimators for one and the 
same LD measure. However, the results for $\eta$ are comparable to those for $D^\prime$ and $Q$. Looking at the results presented in table 2 we can 
summarize the following observations. 
\\ \\
{\bf Observation 1:} For all scenarios and measures, the naive estimator has 
the highest mean square error. 
\\ \\
{\bf Observation 2:} Semi-naive and Bayes estimators perform almost equally well for all measures. However, as expected the Bayes estimator performs best if the defining distribution equals the sampling 
distribution of the tables. The semi-naive estimator is robust against variation of the sampling distribution. 
\\ \\
{\bf Observation 3:} The volume estimator for $D^\prime$ is better than the naive estimator but worse 
than the semi-naive estimators. It is especially worse in case of sampling distribution
${\cal D}\left(\halb\right)$ were small entries of the tables are likely.
\\ \\
{\bf Observation 4:} The volume estimator performs comparable to the semi-naive and Bayes estimator
for both $\eta_1$ and $\eta_\halb$.
\\ \\
Summarising these results, we suggest to use one of the semi-naive estimators to estimate 
all linkage disequilibrium measures considered. The Bayes estimators are not better but are
computationally more expensive. The same holds true for the volume estimators. Moreover, for 
$D^\prime$ the volume estimator is clearly outperformed by the semi-naive estimators 
if the occurrence of small table entries is likely.
 
\subsection*{3.3. Numerical Issues}
The analytic solutions (2.6) and (2.11) cause numerical problems, because of the singularity for $\lambda=1$ in combination with the differences in the numerator. Hence, in the neighbourhood of $\lambda=1$ 
it is useful to replace the analytic formula by the corresponding {\sc Taylor} series. 
After some calculation one finds that
\begin{eqnarray}
\eta_1\left(1+\varepsilon\right)&=&\frac{1}{6}\left(2\varepsilon-\varepsilon^2\right)+{\cal{O}}\left({\varepsilon}^3\right) \\
\eta_\halb\left(1+\varepsilon\right)&=&\frac{1}{\pi^2}\left(2\varepsilon-\varepsilon^2\right)+{\cal{O}}\left({\varepsilon}^3\right)
\end{eqnarray}
where ${\cal O}$ is the first {\sc Landau} order symbol.
\\
The calculation of the Bayes estimator of $\eta$ is also computationally complicate. We suggest to use 
Monte-Carlo integration in combination with a quick sampling tool for Dirichlet distributions. 
The calculation of the volume estimator for $\eta$ is computationally expensive as well if the number of haplotypes is high, since computational effort rises with ${\cal O}\left( N^3\right)$.

Algorithms of all methods have been implemented in the statistical software R \citep{Iha}.
We will provide the scripts upon request. 

\section*{4. Discussion}
\setcounter{chapter}{4}

In this paper we proposed and justified six postulates for a canonical measure of (allelic) association (linkage disequilibrium) intended for application to one-sample two by two contingency tables ${\mathbb T}$: The measure is a mapping of ${\mathbb T}$ to the set of real numbers. It should be zero in case of independence and extremal if one of the entries approaches zero while the marginals are positively bounded. It should reflect the symmetry group of two by two tables and be invariant under certain transformations of the marginals (selection invariant). Their scale should be maximally discriminative for arbitrary tables relative to a calibrating (symmetric) distribution on the manifold of two by two probability tables. \\ 
We proved that there is a unique canonical LD measure for each choice of a calibrating symmetric distribution on ${\mathbb T}$. This calibrating distribution specifies an easy-to-interpret scale essentially based on the fraction of tables exhibiting a less extreme odds ratio than the given one. Although we will use Bayesian and empirical Bayesian considerations in the following in order to motivate the choice of the calibrating distribution, it is only nice but not necessary that the calibrating distribution is a proper Bayesian prior for data at hand.   
\\
The canonical LD measures have maximum entropy relative to their defining calibrating distribution. The principle of maximum entropy classifiers is not new and has been applied to several areas of interest (for example \citet{Nig,Zhu}). However, to our knowledge there is no application of maximum entropy classifiers to the problem of association measures of two by two contingency tables. 
\\
Theoretical and empirical arguments support the choice of ${\cal D}\left(\halb\right)$ as calibrating distribution. ${\cal D}\left(\halb\right)$ is Jeffreys' non-informative prior on ${\mathbb T}$ derived from an information invariance principle \citep{Jef}. ${\cal D}\left(\halb\right)$ induces the uniform distribution on the marginal frequencies and is weakly informative concerning the odds ratio (confer Figure 1). Empirically in our experience, SNP-array data often exhibit a rather uniform distribution of minor allele frequencies when disregarding extremely rare SNPs (confer Figure 6). Consequently, $\eta_\halb$ tends to have a roughly uniform distribution when calculating pair-wise LD in a small region of the genome (confer Figure 7). Thus $\eta_\halb$ can also be interpreted in an empirical Bayesian way as the fraction of tables in the analysed data exhibiting a less extreme odds ratio than the given one. 
\\Hence, for applications in SNP data we recommend the use of $\eta_\halb$. In situations where most tables have less imbalanced marginals, $Q$ (corresponding to $\eta_2$) is a reasonable alternative. 
\\
The popular measures  $D^\prime$ and $r$ are not selection invariant. $D^\prime$ is motivated by a biological model of human evolution and genomic structure which is not in the focus of our biometrical point of view \citep{Mort,Shete}. Selection invariance is particularly important if one wants to compare LD between pairs of SNPs across the genome or across different populations. In this case one needs a measure of association that can be compared between tables with markedly different marginal distributions (allele frequencies).
\\
The measure $r$ (and $M\!I$) is extremal when a diagonal of the table tends to zero. The canonical measure is extremal when there is one zero in the table because an emerging single SNP gives rise to a table with 
one zero. On the other hand, measures which are extremal only for tables with a diagonal zero are pertinent 
when measuring the degree of redundancy between single markers. 
\\
We sharply distinguish between the definition of a LD measure and its estimation. 
For $D^\prime$ the estimation problem has been considered recently and partly by \citet{Seb}. \citet{Lo} investigated jackknife and bootstrap estimators for $D^\prime$.
\\
The usual naive plug-in estimators based on frequencies can lead to unreliable estimates \citep{Chen,Lo,Seb}. Estimation functions based on the computationally expensive volume measures \citep{Chen} were proposed recently as a remedy to this well-known problem.
\\ Here, we investigated four different consistent estimation functions for the measures $\eta$, $D^\prime$, $r$ and $Q$, the naive estimator, the semi-naive estimator, the Bayes estimator and the volume estimator (for $\eta$, $D^\prime$ only) and compared them in an extensive simulation study based on the expected mean square error \citep{Lo}.
\\
We confirmed that volume estimators have better expected mean square error than the naive plug-in estimators.  
In the case of $D^\prime$, volume estimation perform worse than the semi-naive estimator particularly for 
the sampling distribution ${\cal D}\left(\halb\right)$. The reason is that the volume definition for $D^\prime$ is based on tables with fixed marginals. Implicitly the marginals are treated as certain but they are in fact random. In contrast, our volume estimator for $\eta$ treats the marginals as random and its accuracy is reasonable. 
\\
In our study the semi-naive estimator outperforms the naive estimator with respect to accuracy and the volume and Bayesian estimators with respect to computational cost.
\\
In summary we propose a canonical measure $\eta_\halb$ for analysing linkage disequilibrium in the one-sample case.  The canonical measure is uniquely characterised by a set of six biometrical postulates.  It is easy to interpret and can be economically calculated and estimated by the semi-naive estimator using R functions which we will be glad to provide on request.

\newpage 
\section*{Figures and Tables}

\subsection*{Figures}
\begin{psfrags}
\setlength{\parindent}{0em}
\psfrag{A}{\normalsize{${\cal D}\left(5,5,5,5\right)$}}
\psfrag{B}{\normalsize{${\cal D}\left(2,2,2,2\right)$}}
\psfrag{C}{\normalsize{${\cal D}\left(1,1,1,1\right)$}}
\psfrag{D}{\normalsize{${\cal D}\left(\frac{1}{2},\frac{1}{2},\frac{1}{2},\frac{1}{2}\right)$}}
\psfrag{E}{\normalsize{${\cal D}\left(\frac{1}{5},\frac{1}{5},\frac{1}{5},\frac{1}{5}\right)$}}
\psfrag{F}{\normalsize{${\cal D}\left(2,1,\frac{1}{2},\frac{1}{5}\right)$}}
\psfrag{G}{\LARGE{$\ln\lambda$}}
\psfrag{H}{\LARGE{$f\left(\ln\lambda\right)$}}
\hspace{0.2cm}\scalebox{0.65}{\includegraphics[0,0][635,463]{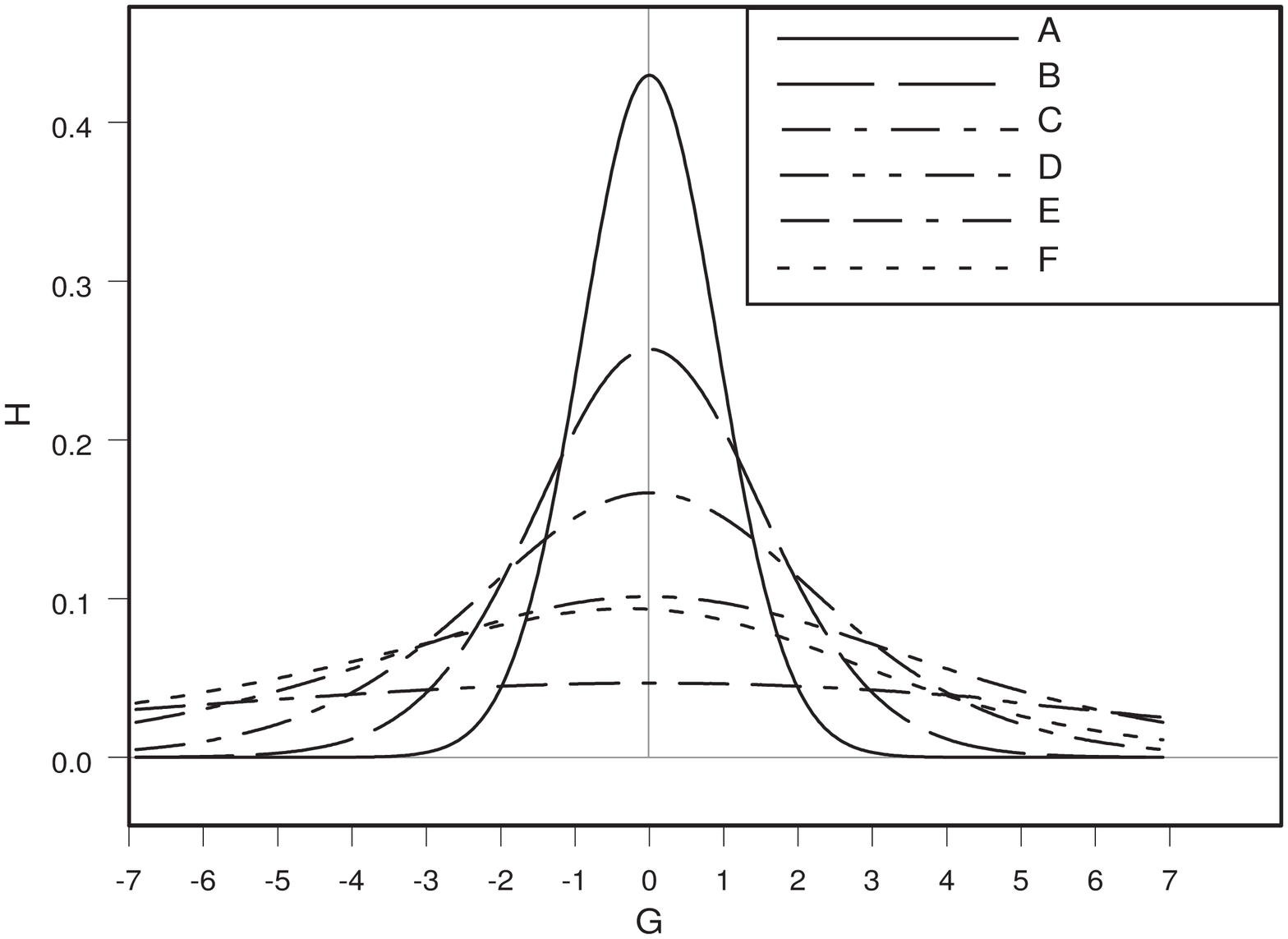}}
\end{psfrags}

\noindent
Figure 1: Density of the log odds ratio under different Dirichlet distributions

\newpage
\setlength{\parindent}{0em}
\psfrag{A}{\Large{$Q$}}
\psfrag{B}{\Large{$\eta_1$}}
\psfrag{C}{\Large{$\eta_\halb$}}
\hspace{0.2cm}\scalebox{0.65}{\includegraphics[0,0][516,170]{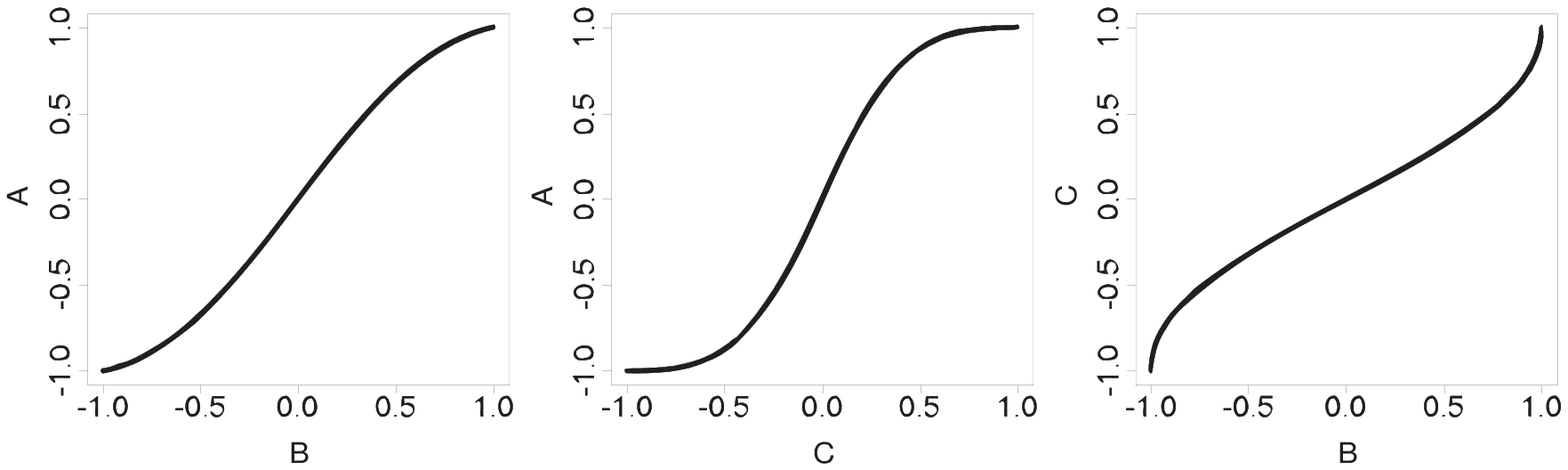}}

\noindent
Figure 2: Relation between odds ratio based measures of LD

\newpage
\setlength{\parindent}{0em}
\psfrag{A}{\Large{$D^\prime$}}
\psfrag{B}{\Large{$Q$}}
\psfrag{D}{\Large{$\eta_1$}}
\psfrag{C}{\Large{$\eta_\halb$}}
\psfrag{E}{\Large{$r$}}
\hspace{0.2cm}\scalebox{0.65}{\includegraphics[0,0][689,170]{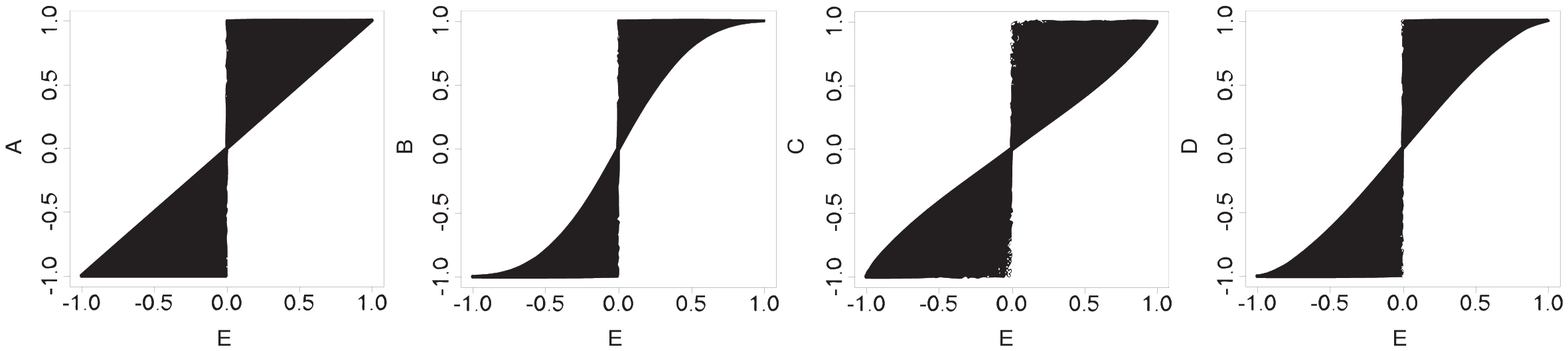}}

\noindent
Figure 3: Correlation of $r$ with other measures of LD (100,000 simulations from 
${\cal D}{\left(\halb\right)}$).

\newpage
\setlength{\parindent}{0em}
\psfrag{A}{\Large{$Q$}}
\psfrag{C}{\Large{$\eta_1$}}
\psfrag{B}{\Large{$\eta_\halb$}}
\psfrag{E}{\Large{$D^\prime$}}
\hspace{0.2cm}\scalebox{0.65}{\includegraphics[0,0][518,170]{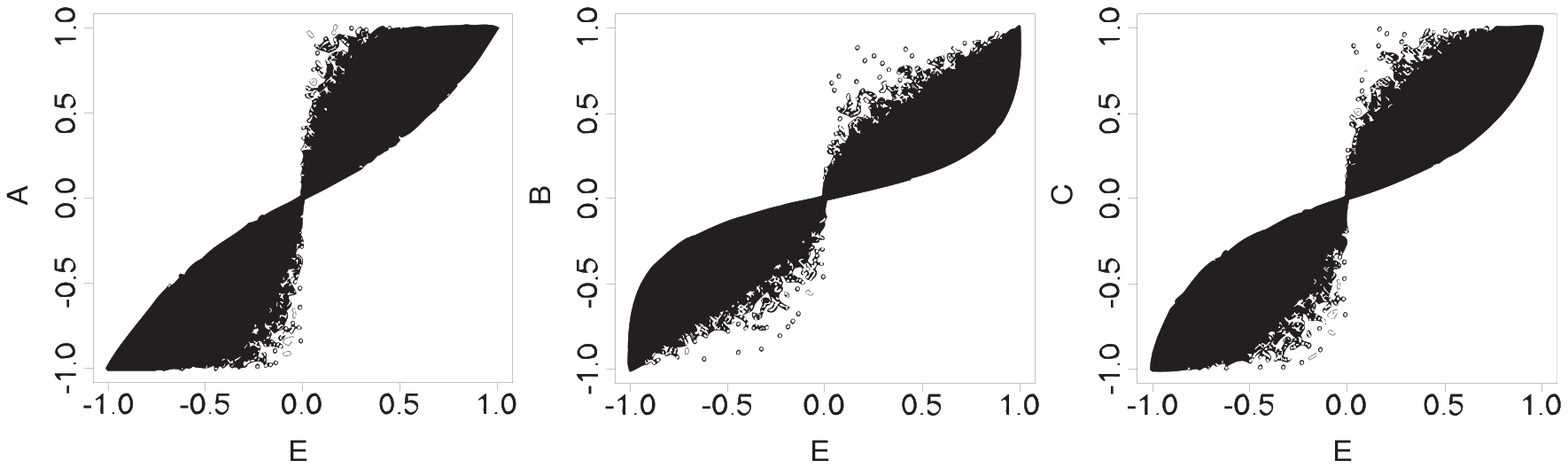}}

\noindent
Figure 4: Correlation of $D^\prime$ with odds ratio based measures of LD (100,000 simulations from ${\cal D}{\left(\halb\right)}$).

\newpage
\begin{psfrags}
\setlength{\parindent}{0em}
\psfrag{A}{\small{$D^\prime$}}
\psfrag{B}{\small{$r$}}
\psfrag{C}{\small{$Q$}}
\psfrag{D}{\small{$\eta_\halb$}} 
\psfrag{E}{\small{\hspace{0.5cm} ${\cal D}\left(\halb\right)$}} 
\psfrag{F}{\small{frequency}}
\psfrag{G}{\small{\hspace{0.5cm} ${\cal D}\left(1\right)$}}
\psfrag{H}{\small{$\eta_1$}}
\hspace{0.2cm}\scalebox{0.65}{\includegraphics[0,0][665,273]{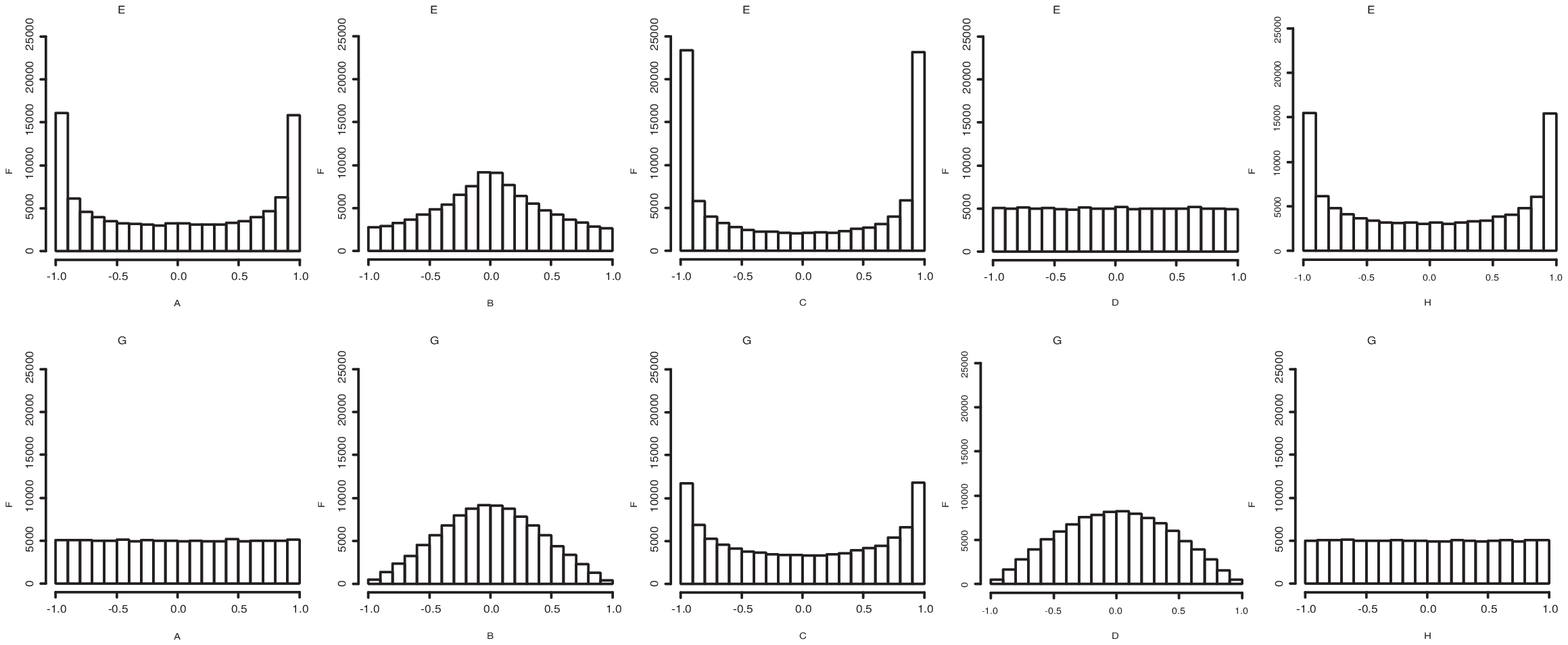}}
\end{psfrags}

\noindent
Figure 5: Distribution of various LD measures assuming a Dirichlet distribution ${\cal D}\left(\halb\right)$ or ${\cal D}\left(1\right)$ on $\mathbb{T}$. By construction, the canonical LD measures $\eta_\halb$ and $\eta_1$ are uniformly distributed for ${\cal D}\left(\halb\right)$ and ${\cal D}\left(1\right)$ respectively. $D^\prime$ is uniformly distributed for ${\cal D}\left(1\right)$. 

\newpage
\setlength{\parindent}{0em}
\hspace{0.2cm}\scalebox{0.65}{\includegraphics[0,0][537,510]{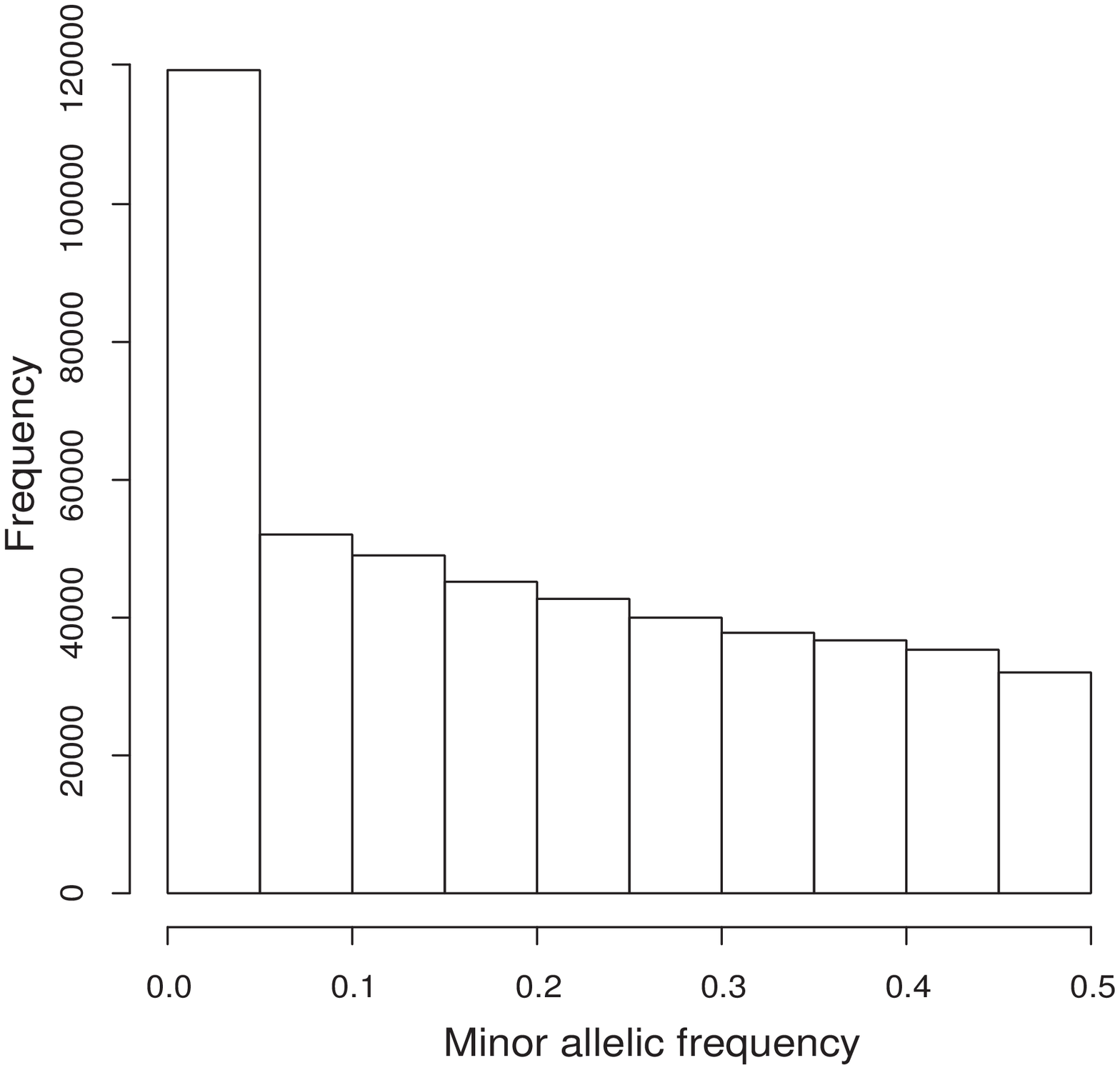}}
\noindent

Figure 6: Distribution of minor allelic frequency in the CEU HapMap sample genotyped with Mapping 500k Affymetrix
chipset \citep{HapMap}. 

\newpage
\setlength{\parindent}{0em}
\psfrag{A}{\normalsize{$Q$}}
\psfrag{B}{\normalsize{$\eta_\frac{1}{2}$}}
\psfrag{F}{\small{frequency}}
\hspace{0.2cm}\scalebox{1}{\includegraphics[0,0][426,182]{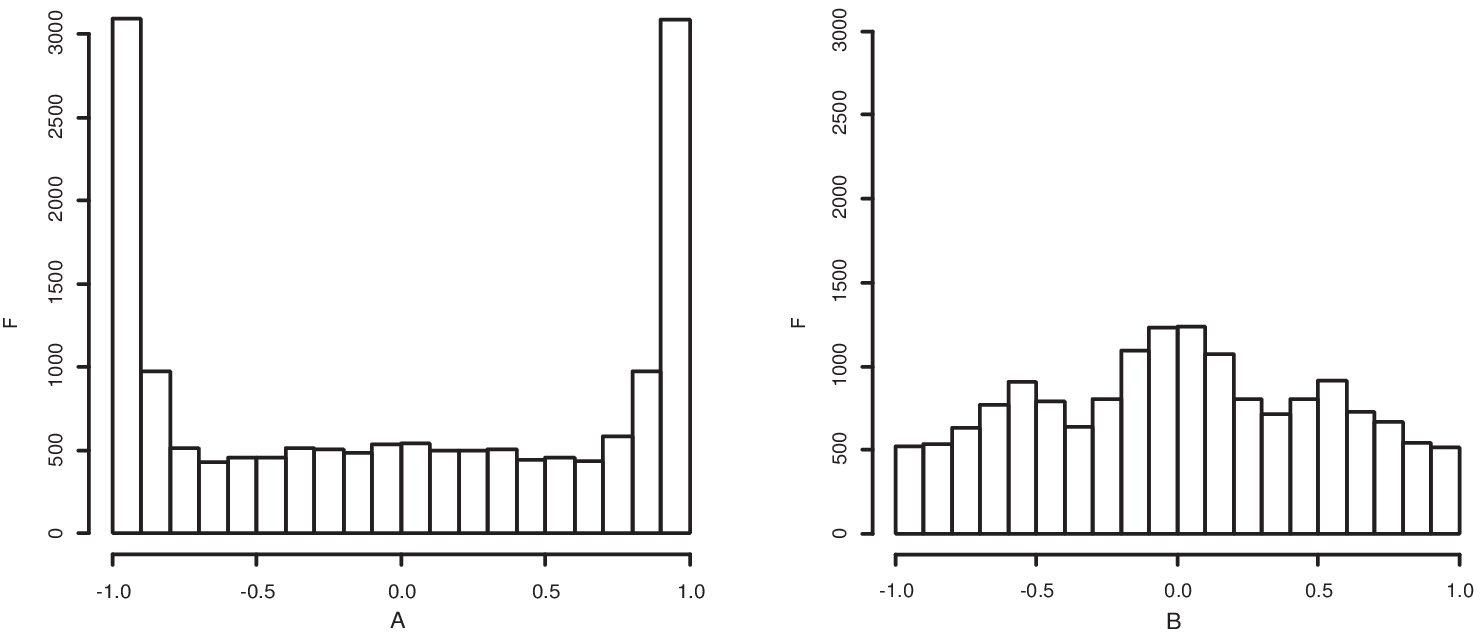}}

\noindent
Figure 7: Distribution of $Q$ and $\eta_\frac{1}{2}$ for all pairs of markers of
chromosome 22 of the CEU HapMap sample. We selected pairs of SNPs with minor allelic frequency greater than 10\% and distances less than 50kb. Semi-naive estimators using $\alpha=\halb$ were calculated. It reveals that $\eta_\halb$ differentiates considerably better between tables than $Q$.

\newpage
\subsection*{Tables}

\begin{tabular}{|r|r|} \hline
Minor marginal frequencies & Kendall's $\tau$ \\ \hline\hline
0\% to 10\% & 0.873 \\ \hline
10\% to 20\% & 0.905 \\ \hline
20\% to 30\% & 0.916 \\ \hline
30\% to 40\% & 0.930 \\ \hline
40\% to 50\% & 0.957 \\ \hline
\end{tabular}
\\ \\ \noindent
{\bf Table 1:} Kendall's correlation coefficient between $D^\prime$ and $\lambda$ for tables with specified marginal frequencies (based on a sample of $N=1,000,000$ drawn from ${\cal D}\left(\halb\right)$). 

\newpage
\noindent
\small{
\begin{tabular}{|r||r|r|r||r|r|r||r|r|r|} \hline
Estimator & \multicolumn{3}{c||}{${\cal D}\left(\halb,\halb,\halb,\halb\right)$} & \multicolumn{3}{c||}{${\cal D}\left(1,1,1,1\right)$} & \multicolumn{3}{c|}{${\cal D}\left(2,2,2,2\right)$} \\ \cline{2-10}
& $N=50$ & $N=100$ & $N=500$ & $N=50$ & $N=100$ & $N=500$ & $N=50$ & $N=100$ & $N=500$ \\ \hline\hline
$\hat{\eta}^{N\!E}_1$              & 0.127 & 0.084 & 0.027 & 0.083 & 0.039 & 0.0064 & 0.050 & 0.022 & 0.0040 \\ \hline
$\hat{\eta}^{S\!N\!E}_1$           & 0.070 & 0.047 & 0.016 & 0.040 & 0.022 & 0.0052 & 0.033 & 0.018 & 0.0038 \\ \hline
$\hat{\eta}^{B\!E}_1$              & 0.063 & 0.042 & 0.014 & 0.040 & 0.022 & 0.0051 & 0.034 & 0.018 & 0.0038 \\ \hline
$\hat{\eta}^{V\!E}_1$              & 0.067 & 0.046 & 0.016 & 0.041 & 0.023 & 0.0052 & 0.037 & 0.019 & 0.0039 \\ \hline\hline
$\hat{\eta}^{N\!E}_\halb$          & 0.140 & 0.093 & 0.0304 & 0.086 & 0.039 & 0.0056 & 0.038 & 0.0134& 0.0019 \\[0.1cm] \hline
$\hat{\eta}^{S\!N\!E}_\halb$       & 0.040 & 0.027 & 0.0095 & 0.022 & 0.012 & 0.0029 & 0.017 & 0.0088& 0.0018 \\[0.1cm] \hline
$\hat{\eta}^{B\!E}_\halb$          & 0.036 & 0.024 & 0.0085 & 0.025 & 0.014 & 0.0031 & 0.020 & 0.0097& 0.0018 \\[0.1cm] \hline
$\hat{\eta}^{V\!E}_\halb$          & 0.040 & 0.027 & 0.0094 & 0.029 & 0.015 & 0.0033 & 0.025 & 0.0116& 0.0020 \\[0.1cm] \hline\hline
$\hat{D^\prime}_{N\!E}$            & 0.115 & 0.074 & 0.023 & 0.081 & 0.039 & 0.0072 & 0.053 & 0.024 & 0.0045 \\ \hline
$\hat{D^\prime}_{S\!N\!E\_ 1}$     & 0.080 & 0.053 & 0.017 & 0.048 & 0.027 & 0.0065 & 0.037 & 0.020 & 0.0043 \\ \hline
$\hat{D^\prime}_{S\!N\!E\_ \halb}$ & 0.067 & 0.045 & 0.015 & 0.050 & 0.028 & 0.0065 & 0.041 & 0.021 & 0.0044 \\[0.1cm] \hline
$\hat{D^\prime}_{B\!E\_ 1}$        & 0.072 & 0.047 & 0.015 & 0.047 & 0.027 & 0.0064 & 0.037 & 0.020 & 0.0044 \\ \hline
$\hat{D^\prime}_{B\!E\_ \halb}$    & 0.064 & 0.042 & 0.014 & 0.051 & 0.028 & 0.0065 & 0.043 & 0.022 & 0.0044 \\[0.1cm] \hline
$\hat{D^\prime}_{V\!E}$            & 0.114 & 0.069 & 0.021 & 0.058 & 0.031 & 0.0067 & 0.038 & 0.020 & 0.0044 \\ \hline\hline
$\hat{r}_{N\!E}$                   & 0.017 & 0.0087& 0.0019& 0.017 & 0.0085& 0.0017 & 0.018 & 0.0090& 0.0018 \\ \hline
$\hat{r}_{S\!N\!E\_ 1}$            & 0.015 & 0.0082& 0.0018& 0.015 & 0.0078& 0.0017 & 0.016 & 0.0084& 0.0018 \\ \hline
$\hat{r}_{S\!N\!E\_ \halb}$        & 0.014 & 0.0078& 0.0018& 0.016 & 0.0080& 0.0017 & 0.017 & 0.0086& 0.0018 \\[0.1cm] \hline
$\hat{r}_{B\!E\_ 1}$               & 0.015 & 0.0082& 0.0018& 0.015 & 0.0078& 0.0017 & 0.016 & 0.0084& 0.0018 \\ \hline
$\hat{r}_{B\!E\_ \halb}$           & 0.014 & 0.0078& 0.0018& 0.015 & 0.0079& 0.0017 & 0.017 & 0.0086& 0.0018 \\[0.1cm] \hline\hline
$\hat{Q}_{N\!E}$                   & 0.135 & 0.090 & 0.029& 0.095 & 0.047 & 0.0085 & 0.071 & 0.034 & 0.0065 \\ \hline
$\hat{Q}_{S\!N\!E\_ 1}$            & 0.087 & 0.059 & 0.020& 0.059 & 0.033 & 0.0076 & 0.055 & 0.030 & 0.0064 \\ \hline
$\hat{Q}_{S\!N\!E\_ \halb}$        & 0.076 & 0.051 & 0.017& 0.061 & 0.034 & 0.0077 & 0.060 & 0.031 & 0.0064 \\[0.1cm] \hline
$\hat{Q}_{B\!E\_ 1}$               & 0.080 & 0.053 & 0.018& 0.058 & 0.033 & 0.0076 & 0.055 & 0.030 & 0.0064 \\ \hline
$\hat{Q}_{B\!E\_ \halb}$           & 0.072 & 0.047 & 0.016& 0.062 & 0.034 & 0.0077 & 0.059 & 0.031 & 0.0064 \\[0.1cm] \hline\hline
\end{tabular}
}
\\ \\ \noindent
\normalsize
{\bf Table 2:} The expected mean square error for different estimators of different LD measures 
based on 100,000 simulations of true tables drawn from the Dirichlet distribution in the columns 
and their realizations with sample size $N$. The estimators are explained in the text.
Except for the canonical measures, we calculate the semi-naive and Bayesian estimators for both 
$\alpha=\halb$ and $\alpha=1$ as well. 

\newpage
\begin{thebibliography}{}
\bibitem[Hapmap()]{HapMap} The HapMap project http://www.hapmap.org/
\bibitem[Chen et al.(2006)]{Chen} {\sc Chen, Y., Lin, C.H.L., Sabatti, C.} (2006). Volume Measures for Linkage Disequilibrium. {\it BMC Genetics} {\bf 7}(54).
\bibitem[Devlin \& Risch(1995)]{Dev} {\sc Devlin, B., Risch, N.} (1995). A Comparison of Linkage Disequilibrium Measures for Fine-Scale Mapping. {\it Genomics} {\bf 29}, 311-322.
\bibitem[Diaconis \& Efron(1985)]{DE} {\sc Diaconis, P., Efron, B.} (1985). Testing for Independence in a Two-Way Table: New Interpretations of the Chi-Square Statistic. {\it The Annals of Statistics} {\bf 13}(3), 845-874.
\bibitem[Edwards(1963)]{Ed} {\sc Edwards, A.W.F.} (1963). The Measure of Association in a 2x2 Table. {\it Journal of the Royal Statistical Society, Series A} {\bf 126}, 108-114.
\bibitem[Gabriel et al.(2002)]{Gabi} {\sc Gabriel, S.B., Schaffner, S.F., Nguyen, H., Moore, J.M., Roy, J., Blumenstiel, B., Higgins, J., DeFelice, M., Lochner, A., Faggart, M., Liu-Cordero, S.N., Rotimi, C., Adeyemo, A., Cooper, R., Ward, R., Lander, E.S., Daly, M.J., Altshuler, D.} (2002). The structure of haplotype blocks in the human genome.
{\it Science} {\bf 296}(5576), 2225-2229.
\bibitem[Geisser(1984)]{Geis} {\sc Geisser, S.} (1984). On Prior Distributions for Binary Trials. {\it The American Statistican} {\bf 38}(4), 244-251.
\bibitem[Hartung(1991)]{Hart} {\sc Hartung, J.} (1991). {\it Statistik: Lehr- und Handbuch der angewandten Statistik}. Munich: R. Oldenbourg Verlag GmbH
\bibitem[Hedrick(1987)]{Hed} {\sc Hedrick, P.W.} (1987). Gametic Disequilibrium Measures: Proceed with Caution. {\it Genetics} {\bf 117}, 331-341.
\bibitem[Hill \& Robertson(1968)]{HR} {\sc Hill, W.G., Robertson, A.} (1968). Linkage Disequilibrium in Finite Populations. {\it Theoretical and Applied Genetics} {\bf 38}, 226-231.
\bibitem[Hotelling(1939)]{Hot} {\sc Hotelling, H.} (1939). Tubes and Spheres in n-Spaces, and a Class of Statistical Problems. {\it American Journal of Mathematics} {\bf 61}, 440-460.
\bibitem[Ihaka \& Gentleman(1996)]{Iha} {\sc Ihaka, R., Gentleman, R.} (1996). R a language for data analysis and graphics. {\it Journal of computational and graphical statistics} {\bf 5}, 299-314. 
\bibitem[Jeffreys(1961)]{Jef} {\sc Jeffreys, H.} (1961). {\it Theory of Probability}. London: Oxford University Press, Inc.
\bibitem[Koelbig()]{Koel} {\sc Koelbig, K.S.} {\it Collected Algorithms from CACM} {\bf 327}-P 1-0.
\bibitem[Lewontin(1963)]{Lew} {\sc Lewontin, R.C.} (1963). The Interaction of Selection and Linkage. I. General Considerations; Heterotic Models. {\it Genetics} {\bf 49}, 49-67.
\bibitem[Li et al.(2008)]{Li} {\sc Li, Y.M., Xiang, Y., Sun, Z.Q.} (2008). An Entropy-Based Measure for QTL Mapping Using Extreme Samples of Population. {\it Human Heredity} {\bf 65}, 121-128.
\bibitem[Lo(1991)]{Lo} {\sc Lo, S.K.} (1991). On the analysis and application of measures of linkage disequilibrium. {\it Australian Journal of Statistics} {\bf 33}(3), 249-259.
\bibitem[Maximom(2003)]{Max} {\sc Maximom, L.C.} (2003). The dilogarithm function for complex argument. {\it Proceedings of the Royal Society of London A} {\bf 459}, 2807-2819.
\bibitem[Morton et al.(2001)]{Mort} {\sc Morton, N.E., Zhang, W., Taillon-Miller, P., Ennis, S., Kwok, P.Y., Collins, A.} (2001). The optimal measure of allelic association. {\it Proceedings of the National Academy of Science} {\bf 98}(9), 5217-5221. 
\bibitem[Mueller(2004)]{Mul} {\sc Mueller, J.C.} (2004). Linkage Disequilibrium for Different Scales and Applications. {\it Briefings in Bioinformatics} {\bf 5}(4), 355-364.
\bibitem[Nigam et al.(1999)]{Nig} {\sc Nigam, K., Lafferty, J., McCallum, A.} (1999). Using Maximum Entropy for Text Classification. {\it IJCAI-99 Workshop on Machine Learning for Information Filtering}, 61-67.
\bibitem[Pritchard \& Przeworski(2001)]{Prit} {\sc Pritchard, J.K., Przeworski, M.} (2001). Linkage disequilibrium in Humans: Models and Data. {\it American Journal of Human Genetics} {\bf 69}, 1-14.
\bibitem[Schulze et al.(2004)]{Schu} {\sc Schulze, T.G., Zhang, K., Chen, Y., Akula, N., Sun, F., McMahon, F.J.} (2004). Defining haplotype blocks and tag single-nucleotide polymorphisms in the human genome. {\it Human Molecular Genetics} {\bf 13}(3), 335-342.
\bibitem[Sebastiani \& Abad-Grau(2007)]{Seb} {\sc Sebastiani, P., Abad-Grau, M.M.} (2007). Bayesian estimates of linkage disequilibrium. {\it BMC Genetics} {\bf 8}(36) doi:10.1186/1471-2156-8-36
\bibitem[Service et al.(2006)]{Serv} {\sc Service, S., DeYoung, J., Karayiorgou, M., Roos, J.L., Pretorious, H., Bedoya, G., Ospina, J., Ruiz-Linares, A., Macedo, A., Palha, J.A., Heutink, P., Aulchenko, Y., Oostra, B., vanDuijn, C., Jarvelin, M.R., Varilo, T., Peddle, L., Rahman, P., Piras, G., Monne, M., Murray, S., Galver, L., Peltonen, L., Sabatti, C., Collins, A., Freimer, N.} (2006). Magnitude and Distribution of Linkage Disequilibrium in Population Isolates and Implications for Genome-wide Association Studies. {\it Nature genetics} {\bf 38}(5), 556-560.
\bibitem[Shete(2003)]{Shete} {\sc Shete, S.} (2003). A Note on the Optimal Measure of Allelic association. {\it Annals of Human Genetics} {\bf 67}, 189-191.
\bibitem[Tear et al.(2002)]{Tear} {\sc Teare, M.D., Dunning, A.M., Durocher, F., Rennart, G., Easton, D.F.} (2002). Sampling distribution of summary linkage disequilibrium measures. {\it Annals of Human Genetics} {\bf 66}, 223-233.
\bibitem[Thomas(2004)]{T} {\sc Thomas, D.C.} (2004). {\it Statistical Methods in Genetic Epidemiology}. New York, NY: Oxford University Press, Inc.  
\bibitem[Walley(1996)]{Wal} {\sc Walley, P.} (1996). Inference from Multinomial Data: Learning About a Bag of Marbles. {\it Journal of the Royal Statistical Society, Series B} {\bf 58}(1), 3-57.
\bibitem[Weaver \& Shannon(1963)]{Shan} {\sc Weaver, W., Shannon, C.E.} (1963). {\it The Mathematical Theory of Communication.} Urbana, IL: University of Illinois Press
\bibitem[Weir(1996)]{Weir} {\sc Weir, B.S.} (1996). {\it Genetic Data Analysis II.} Sunderland, MA: Sinauer Associates, Inc.
\bibitem[Yule(1900)]{Yule} {\sc Yule, G.U.} (1900). On the Association of Attributes in Statistics. {\it Philosophical Transactions of the Royal Society of London, Series A 1900} {\bf 194}, 269-274.
\bibitem[Zhang et al.(2002)]{ZDC} {\sc Zhang, K., Deng, M., Chen, T., Waterman, M.S., Sun, F.} (2002). A dynamic programming algorithm for haplotype block partitioning. {\it Proceedings of the National Academy of Science} {\bf 99}(11), 7335-7339.
\bibitem[Zhu et al.(2005)]{Zhu} {\sc Zhu, S., Ji, X., Xu, W., Gong, Y.} (2005). Multi-labelled Classification Using Maximum Entropy Method. {\it Proceedings of the 28th annual international ACM SIGIR conference on Research and Development in Information Retrieval}, 274-281.
\end{thebibliography}
\end{document}